\documentclass[a4paper,12pt]{article}
\usepackage[dvips]{graphicx}

\newcommand{\VCT}[1]{\mbox{\boldmath$#1$}}

\begin{document}

\normalsize

\onecolumn

\setlength{\baselineskip}{17.5pt}


\vspace*{0.60in}

\begin{center}
{\large\bf 
Multi-ion-species effects on magnetosonic waves and energy transfer 
in thermal equilibrium plasmas
}

\vspace*{0.30in}


   TOIDA Mieko, YOSHIYA Takashi,  and OHSAWA Yukiharu 


{\it 
   Department of Physics, Nagoya University, 
        Nagoya, 464-8602, Japan  \\
}


{\it 
   e-mail: toida@phys.nagoya-u.ac.jp
}

\end{center}

\vspace*{0.00in}


\noindent
{\bf Abstract}
\setlength{\baselineskip}{4.30mm}

\hspace*{12pt} \small{
Magnetosonic waves propagating perpendicular to an external magnetic field are studied with attention to the effect of multiple ion species. First, power spectra of magnetic field fluctuations and autocorrelation functions in thermal equilibrium plasmas are numerically obtained. In a multi-ion-species plasma, besides $\omega \simeq kv_{\rm A}$ mode, numerous waves are present near many different ion cyclotron frequencies. The autocorrelation function of the quasi-mode consisting of these waves is not recovered to its initial value, owing to the phase mixing of these waves.  Next, with particle simulations, evolution of a macroscopic perpendicular disturbance is investigated. In a multi-ion-species plasma, this disturbance is damped. The energy is transferred to from the magnetic field to the ions.}
\vspace*{0.20in}



\vspace*{0.30in}


\setlength{\baselineskip}{4.4mm}


\noindent
{\bf 1. Introduction}

The presence of multiple ion species introduces many interesting effects on magnetosonic waves [1-7]. For instance, in a two-ion-species plasma, the magnetosonic wave is split into two modes. Nonlinear pulses of these modes are damped, even when they propagate perpendicular to the magnetic field [5-7]. The damping is due to energy transfer from the pulse to heavy ions [8,9]. Periodic waves are not damped even in this case. However, the collective behavior of these waves in a multi-ion-species plasma would be different from that in a single-ion-species plasma. 

Recently, a study has been made on collective behavior of ion Bernstein waves in thermal-equilibrium plasmas with multiple ion species [10]. Each perpendicular ion Bernstein wave with $\omega \simeq n\Omega_{\rm i}$ is undamped in a collisionless plasma [11], where $\Omega_i$ is the ion cyclotron frequency and $n$ is the integer. In a single-ion-species plasma, the autocorrelation function of the quasi-mode consisting of these waves shows periodic behavior with time period $2 \pi/\Omega_i$. On the other hand, in a multi-ion-species plasma, the autocorrelation function is initially damped and is not recovered. This is caused by the phase mixing of numerous waves excited at the harmonics of many different ion cyclotron frequencies.  This damping mechanism could be important in space plasmas where many ion species exist, with each species having many different ionic charge states.

In this paper, we study perpendicular magnetosonic waves in thermal-equilibrium, multi-ion-species plasmas where each particle species has its own Maxwellian velocity distribution.  We assume that all the ion species have an equal temperature, while electrons can have a different temperature because relaxation time between electrons and ions via collisions is very long. 

In Sec. 2, we numerically calculate power spectra and autocorrelation functions of magnetic field fluctuations due to the magnetosonic waves. In a single-ion-species plasma, the autocorrelation function is not damped, because the wave with $\omega \simeq kv_{\rm A}$ is dominant mode. Here, $v_{\rm A}$ is the Alfv\'en speed and $k$ is the perpendicular wavenumber. On the other hand, in a multi-ion-species plasma, besides this mode, numerous waves are present near many different ion cyclotron frequencies. Owing to the phase mixing of these waves, the autocorrelation function does not return its initial value. In Sec. 3, evolution of a macroscopic disturbance and associated energy transport are studied by particle simulations. In a multi-ion-species plasma, the macroscopic disturbance is damped, and the energy is transferred from the magnetic field to the ions.

\vspace*{0.20in}


\noindent
{\bf 2. Numerical Calculation}

We consider extraordinary waves propagating perpendicular to an external magnetic field with frequensies smaller than lower hybrid frequency; we call these waves as magnetosonic waves.  The dispersion relations of the magnetosonic waves are given by 
\begin{equation}
D_{\rm ms} \equiv \varepsilon_{xy}^2/\varepsilon_{xx}  
+ \varepsilon_{yy} - c^2k^2/\omega^2 = 0, 
\label{eq:disp}
\end{equation} 
where $c$ is the light speed, and $\varepsilon_{xx}, \varepsilon_{xy}$, and 
$\varepsilon_{yy}$ are defined as
\begin{equation}
\varepsilon_{xx} = 1 - \sum_j \sum_n \frac{ \omega_{pj}^2}{\omega(\omega - n\Omega_j)} \frac{n^2 }{\mu_j} \Gamma_n(\mu_j),
\label{eq:exx}
\end{equation}
\begin{equation}
\varepsilon_{xy} = - i \sum_j \sum_n \frac{\omega_{pj}^2}{\omega(\omega - n\Omega_j)} n \Gamma'_n(\mu_j), 
\label{eq:exy}
\end{equation}
\begin{equation}
\varepsilon_{yy} = \varepsilon_{xx}+  \sum_j \sum_n \frac{2\omega_{pj}^2}{\omega(\omega - n\Omega_j)} \mu_j \Gamma'_n(\mu_j). 
\label{eq:eyy}
\end{equation} 
Here, the subscript $j$ refers to electrons (e) or ion species (H, He, C, $\cdots$), $\Omega_j$ is the cyclotron frequency, and $\omega_{{\rm p}j}$ is the plasma frequency. Also, $\Gamma_{n} (\mu_j)$ = $I_{n}(\mu_j) \exp (-\mu_j)$, where $I_{n}$ is the modified Bessel function of the $n$th order, and $\mu_j = k^2 \rho_j^2$ with $\rho_j$ the gyro-radius.

The fluctuation spectrum of magnetic fields due to the magnetosonic waves in a spatially homogeneous, thermal equilibrium plasma is written as
\begin{equation}
\frac{|B_{k,\omega}|^2}{8 \pi} = \sum_n P(\omega) \delta(\omega - \omega_{n}),
\label{eq:b-fluct}
\end{equation}
with 
\begin{equation}
P(\omega) = \frac{\pi k_{\rm B} T}
{\displaystyle {\left. \omega \frac{\partial}{\partial \omega} D_{\rm ms} (k,\omega) \right|_{\omega=\omega_{n}}}}
\frac{\varepsilon_{xy}^2}{\varepsilon_{xx}}
\frac{c^2 k^2}{\omega^2},
\label{eq:p(b)}
\end{equation}
where $k_{\rm B}$ is the Boltzmann constant, and $\omega_{n}$ is the roots of the dispersion relation $D_{\rm ms}$=0. 

In a multi-ion-species plasma with H being major ions, there are three kinds of waves in the long wavelength region, $\mu_i \ll 1$; 
\begin{equation}
\omega \simeq kv_{\rm A},
\label{eq:kva}
\end{equation}
\begin{equation}
\omega \simeq \Omega_s + \frac{\omega_{{\rm p}s}^2 \Omega_{\rm H}}{\omega_{\rm pH}^2 \Omega_s} (\Omega_{\rm H} - \Omega_s),
\label{eq:wh0}
\end{equation}
and
\begin{equation}
\omega \simeq n\Omega_i. 
\label{eq:kinetic}
\end{equation}
Here, the subscript $s$ in Eq. (\ref{eq:wh0}) refers to heavy ion species (He, C, O, $\cdots$). The waves with Eqs. (\ref{eq:kva}) and (\ref{eq:wh0}) exist even in a cold plasma [4], while the mode with Eq. (\ref{eq:wh0}) is caused by ion kinetic effects [12]. 

For a given wavenumber $k$, there are many waves with different frequencies.  Autocorrelation function of the quasi-mode consisting of these waves is obtained from $P(\omega)$ though the Fourier transformation in $\omega$ as
\begin{equation}
C_k (\tau) = \int_{- \infty}^{\infty} |B_{k,\omega}|^2 \exp(-i \omega \tau) d\omega.
\label{eq:auto}
\end{equation}
We pay attention to how values of $|C_k(\tau)|$ are reduced by the presence of multiple ion species. The reduction of $|C_k(\tau)|$ indicates that energy transport can be enhanced [13].  

We numerically calculate specific values of $P(\omega)$ and $C_k(\tau)$ for three different plasmas and compare them. [In the calculation, we retain the terms from $n=-$10 to 10 for the ions and the $n=0$ and 1 terms for the electrons in Eqs. (\ref{eq:exx}), (\ref{eq:exy}), and (\ref{eq:eyy}).]  The three plasmas that we consider are single-ion (H$^{+}$), three ion (H$^{+}$, He$^{+2}$, C$^{+5}$) and six-ion (H$^{+}$, He$^{+2}$, C$^{+6}$, O$^{+6}$, Si$^{+9}$, and Fe$^{+13}$) species plasmas. The cyclotron frequencies of these ions normalized to $\Omega_{\rm H}$ are taken to be $\Omega_{\rm He}=0.5$, $\Omega_{\rm C}=0.417$, $\Omega_{\rm O}=0.375$, $\Omega_{\rm Si}=0.321$, and $\Omega_{\rm Fe}=0.232$. The densities of the ions normalized to $n_{\rm H}$ are $n_{\rm He}=0.1$, $n_{\rm C}=n_{\rm O}=0.01$, and $n_{\rm Si}=n_{\rm Fe}=0.005$. The magnetic field strength is $|\Omega_{\rm e}|/\omega_{\rm pe}=1$. The plasma beta value is $\beta = 0.0625$.  

\begin{figure}[htbp]
  \centering
  \includegraphics[width=8cm]{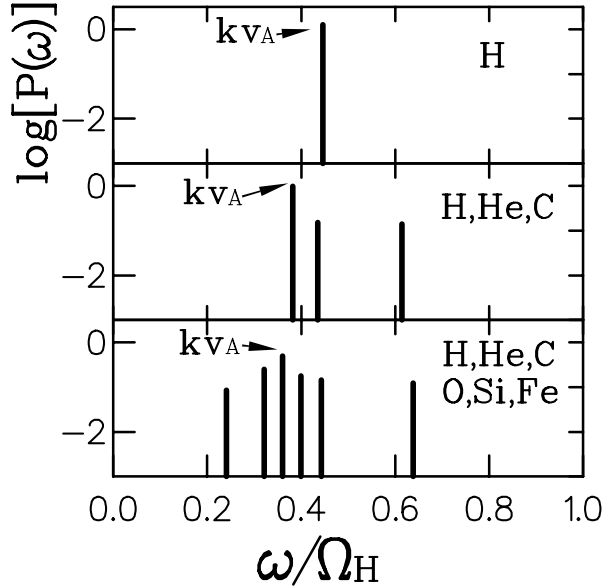}
  \caption{Power spectra of magnetic field fluctuations with $k \rho_{\rm H} = 0.1$  in three different plasmas.}
  \label{fig:n-pw}
\end{figure} 

Figure 1 shows power spectra of the mode with $k \rho_{\rm H} = 0.1$ in the single-, three, and six-ion-species plasmas, where $P(\omega)$ is normalized to $\pi k_B T$. In the single-ion-species plasma, the wave with $\omega \simeq kv_{\rm A}$ ($\simeq 0.42\Omega_{\rm H}$) is the dominant mode. Even though there are waves with $\omega \simeq n\Omega_{\rm H}$, their amplitudes are quite small. In the three-ion-species plasma, besides the $\omega \simeq kv_{\rm A}$ $(\simeq 0.39 \Omega_{\rm H}$) mode, the waves near $\Omega_{\rm He}$ and $\Omega_{\rm C}$ are present; these frequencies are given by Eq. (\ref{eq:wh0}) with $s=$ He or C. The amplitudes of the waves with $\omega \simeq n\Omega_{\rm H}, n\Omega_{\rm He}$, and $n\Omega_{\rm C}$ are much smaller. In the six-ion-species plasma, the waves near $\Omega_{\rm O}$, $\Omega_{\rm Si}$, and $\Omega_{\rm Fe}$ also exist. Their amplitudes are not small, although the abundances of the heavy ions are very small. 

\begin{figure}[htbp]
  \centering
  \includegraphics[width=7cm]{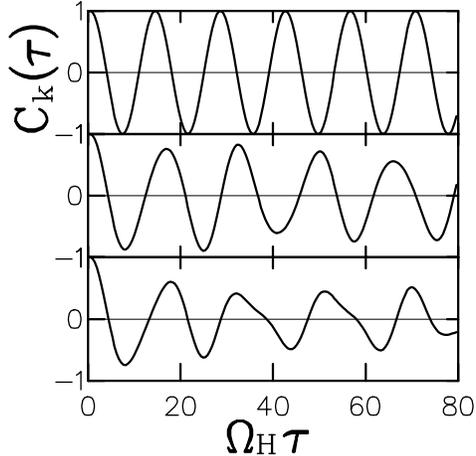}
  \caption{Autocorrelation functions of the fluctuations with $k \rho_{\rm H}=0.1$ in the same plasmas as shown in Fig. 1}
  \label{fig:n-auto}
\end{figure} 

Figure 2 shows autocorrelation functions normalized to their initial values $C_k(0)$.  In the single-ion-species plasma, $C_{\rm k}(\tau)$ oscillates with the period $2 \pi/(kv_{\rm A})$ and is undamped. In the three- and six-ion-species plasmas, $C_{\rm k}(\tau)$'s do not return to their initial values till the end of the calculation. As the number of ion species increases, the amplitude of the oscillation decreases more quickly, owing to the phase mixing of more waves.  

If effects of collisions are entirely neglected, $C_k(\tau)$ returns to its initial value at the time of the least common multiple of all the wave periods. However, this time is extremely long in space plasmas where the number of ion species is very large (moreover, each ion species has many different ionic charge states) and numerous waves exist. On such a long time scale, the effects of collisions must be important, which reduces $|C_k(\tau)|$. Accordingly, $C_k(\tau)$ would not be recovered and would keep smaller values than in a single-ion-species plasma. 

\vspace*{0.20in}


\noindent
{\bf 3. Particle Simulations}

By means of a one-dimensional (one space and three velocity components), electromagnetic particle code with full ion and electron dynamics, we study collective behavior of magnetosonic waves in a multi-ion-species plasma. The system size is $L_x = 512 \Delta_g$, where $\Delta_g$ is the grid spacing and is equal to the electron Debye length.  We use periodic boundary conditions. The external magnetic field is in the $z$ direction, and its strength is $|\Omega_{\rm e}|/\omega_{\rm pe}=4.0$. The total number of electrons is $N_{\rm e}=262,144$. The plasma $\beta$ value is $\beta = 0.03$.

We simulate single-ion (a) and four-ion (a, b, c, and d) species plasmas. We choose the mass ratios as $m_{\rm a}/m_{\rm e} = 50$, $m_{\rm b}/m_{\rm a} = \sqrt{3}$, $m_{\rm c}/m_{\rm a}= 2$, and $m_{\rm d}/m_{\rm a}= \sqrt{5}$. In order to see the effect of multiple ion species with a small number of ion species, we have taken the irrational ion mass rations for b and d ions. The charges are the same, $q_{\rm a} =q_{\rm b} = q_{\rm c} =q_{\rm d} = |q_{\rm e}|$. The ion densities are set to be $n_{\rm b}=n_{\rm c}=n_{\rm d}=0.2 n_{\rm a}$. 

Firstly, we observed that autocorrelation functions of fluctuations propagatig perpendicular to the external magnetic field are not recovered in the four-ion-species plasma. Next, as a initial condition, we set the magnetic field to have a finite amplitude disturbance with a monochromatic cosine profile, 
$
\delta B_{z}(x)/B_{0}  = 0.02 \cos (k_0 x), 
$
where $B_{0}$ is the external magnetic field and $k_0 \rho_{\rm a} = 0.01$($k_0 v_{\rm A}=0.68\Omega_{\rm a}$). We then study evolution of its disturbance and associated energy transport. Initially, all the ion species have equal temperature; the electron-to-ion temperature ratio is chosen to be $T_{i}/T_{\rm e}=0.1$.

\begin{figure}[htbp]
  \centering
  \includegraphics[width=8cm]{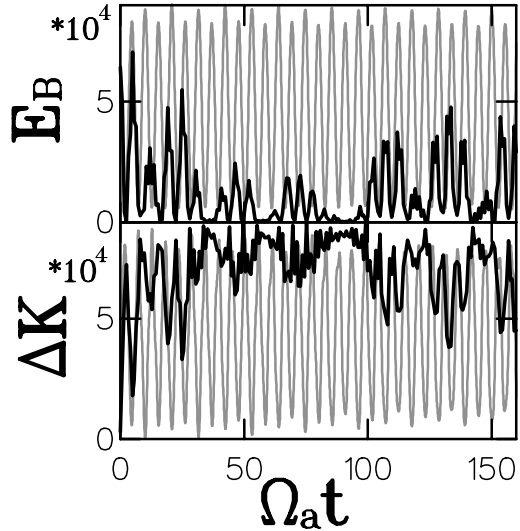}
  \caption{Time variations of total magnetic-field energies and ion kinetic energies. The thin and thick lines represent single- and four-ion-species plasmas, respectively.}
  \label{fig:sm-Bz}
\end{figure} 

Figure 3 shows time variations of total magnetic-field energy $E_B$ and ion energy $K - K_0$, where $K$ is the total energy of all the ions and $K_0$ is the initial one. The energies are normalized to $m_{\rm e} v_{\rm Te}^2$. The thin and thick lines denote energies in the single- and four-ion-species plasmas, respectively. In the single-ion-species plasma, the magnetic field energy and ion kinetic energy oscillate with period $\pi/(k_0 v_{\rm A})$. On the other hand, in the four-ion-species plasma, the magnetic field energy is reduced and does not return to its initial value. The ion energy is rapidly increased and then keeps large values. 

\begin{figure}[htbp]
  \centering
  \includegraphics[width=8cm]{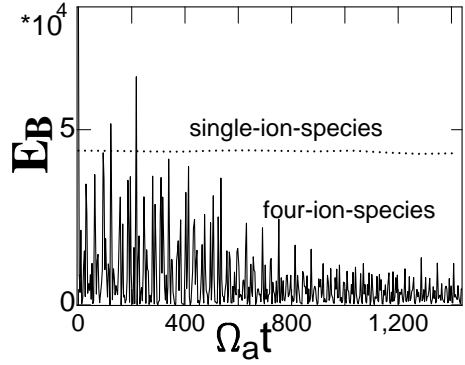}
  \caption{Long time variations of total magnetic-field energies. The solid line and dashed lines represent the energies in the four- and single-ion-species plasmas, respectively. For the single-ion-species plasma, the values averaged over the time period $t = 2 \pi/k_0 v_{\rm A}$ are plotted.}
  \label{fig:eb-long}
\end{figure}

Figure 4 shows long time variations of total magnetic field energies in the single- and four-ion-species plasmas. (For the single-ion-species plasma, the values averaged over the time period $t = 2 \pi/(k_0 v_{\rm A})$ are plotted.)  The magnetic field energy in the single-ion-species plasma keeps almost constant even for the long period. On the other hand, in the four-ion-species plasma, the magnetic-field energy is eventually damped.

Figure 5 shows time variations of ion kinetic energies, which is defined as 
\begin{equation}
KE =  \sum_i \int d \VCT{x} 
\int d \VCT{v} m_i f_{i}(\VCT{x},\VCT{v}) (\VCT{v} - <\VCT{v}_i(\VCT{x})>)^2,
\label{eq:Ti}
\end{equation}
with $<\VCT{v}_i(\VCT{x})>$  the fluid velocity at position $\VCT{x}$, and $KE(0)$ is the initial value of $KE$. In the four-ion-species plasma, the ion kinetic energy is increased. Evidently, the presence of the multiple ion species can enhance the energy dissipation.

\begin{figure}[htbp]
  \centering
  \includegraphics[width=8cm]{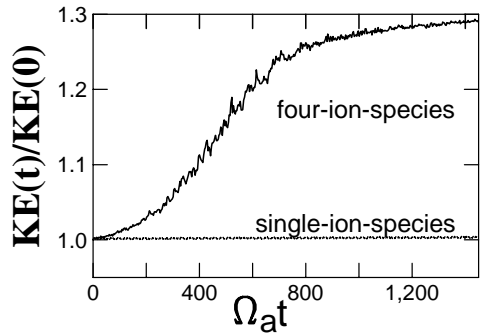}
  \caption{Long time variations of ion kinetic energies in the single- and four-ion-species plasmas.}
  \label{fig:eb-long}
\end{figure}

\vspace*{0.2in}
\noindent
{\bf 4. Summary}

We have studied collective behavior of perpendicular magnetosonic waves in multi-ion-species plasmas. We have numerically shown that the autocorrelation functions in a thermal-equilibrium plasma are not recovered, because, in addition to the $\omega \simeq k v_{\rm A}$ mode, many waves exist near many different cyclotron frequencies. Furthermore, we have shown with particle simulations, that the macroscopic disturbance is also damped in a multi-ion-species plasma, and that the energy is transferred from the magnetic field to the ions. We have not yet understood the transfer mechanism. To do this, we will further investigate with particle simulations, for example, how the energies depend on the initial conditions or on the ion species. 

\vspace*{0.20in}


\noindent
{\bf References}

\noindent
[1] S. J. Buchsbaum, Phys. Fluids {\bf 3}, 418 (1960).

\noindent
[2] A. B. Mikhailovskii and A. I. Smolyakov, Sov. Phys. JETP 
{\bf 61}, 109 (1985).

\noindent
[3] U. Motschmann, K. Sauer, T. Roatsch, and J. F. Mckenzie, 
J. Geophys. Res. {\bf 96}, 13841 (1991).

\noindent
[4] M. Toida and Y. Ohsawa, J. Phys. Soc. Jpn. {\bf 63}, 573 (1994).

\noindent
[5]  D. Dogen, M. Toida, and Y. Ohsawa, 
Phys. Plasmas {\bf 5}, 1298 (1998).

\noindent
[6] M. Toida, D. Dogen, and Y. Ohsawa, 
J. Phys. Soc. Jpn. {\bf 68}, 2157 (1999).

\noindent
[7] S. Irie and Y. Ohsawa, 
Phys. Plasms {\bf 10}, 1253 (2003).

\noindent
[8] M. Toida and Y. Ohsawa, 
J. Phys. Soc. Jpn. {\bf 64}, 2036 (1995).

\noindent
[9] M. Toida and Y. Ohsawa, 
Solar Physics {\bf 171}, 161 (1997).

\noindent
[10] M. Toida, T. Suzuki, and Y. Ohsawa, 
J. Plasma Fusion Res. {\bf 79}, 549 (2003).

\noindent
[11] I. B. Bernstein, 
Phys. Rev. {\bf 109}, 10 (1958).

\noindent
[12] T. D. Kaladze, D. G. Lominadze, and K. N. Stepanov, Sov. Phys. JETP 
{\bf 7}, 196 (1972)

\noindent
[13] T. Kamimura, T. Wagner, and J. M. Dawson, Physics Fluids {\bf 21}, 1151 (1978).


\end{document}